\documentclass[longbibliography,twocolumn,amsmath,amssymb,aps]{revtex4-1}
\usepackage{graphicx,color,SIunits,bm}

\begin{document}

\title{A universal curve of optimum thermoelectric figures of merit \\
  for bulk and low-dimensional semiconductors}
 
\author{Nguyen T. Hung}
\email{nguyen@flex.phys.tohoku.ac.jp}

\author{Ahmad R. T. Nugraha}
\author{Riichiro Saito}

\affiliation{Department of Physics, Tohoku University, Sendai
  980-8578, Japan}

\begin{abstract}
  Analytical formulas for thermoelectric figure of merit and power
  factor are derived based on the one-band model.  We find that there
  is a direct relationship between the \emph{optimum} figures of merit
  and the \emph{optimum} power factors of semiconductors despite of
  the fact that the two quantities are generally given by different
  values of chemical potentials.  By introducing a dimensionless
  parameter consisting of optimum power factor and lattice thermal
  conductivity (without electronic thermal conductivity), it is
  possible to unify optimum figures of merit of both bulk and
  low-dimensional semiconductors into a single universal curve that
  covers lots of materials with different dimensionalities.
\end{abstract}

\maketitle

\section{Introduction}

Most of electrical energy that we consume in daily life comes from thermal
processes, such as heat engines in cars and power plants, in which
more than half of the energy is wasted in form of
heat~\cite{bell2008cooling}.  Research on thermoelectricity for
recovering this waste heat, i.e. to convert the waste heat directly
into electric energy, is thus of great
interest~\cite{bell2008cooling,heremans2013thermoelectrics}.  A good
thermoelectric (TE) material is characterized by how efficient the
electricity can be obtained for a given heat source, in which the
thermoelectric figure of merit $ZT=S^2\sigma\kappa^{-1}T$ is usually
evaluated, where $S$, $\sigma$, $\kappa$, and $T$ are the Seebeck
coefficient, the electrical conductivity, the thermal conductivity,
and the average absolute temperature, respectively.  It is well-known
that obtaining the optimum $ZT$ (or shortly $ZT_{\rm opt}$) for a
certain TE material, where $ZT_{\rm opt}$ is defined as the maximum
value of $ZT$ as a function of the chemical potential, is often
complicated by the interdependence of $S$, $\sigma$, and
$\kappa$~\cite{vining2009inconvenient}.  Therefore, finding the best
material to obtain as large $ZT_{\rm opt}$ as possible has been a
great challenge for many years.  As one strategy, using
low-dimensional semiconductors with a large density of states at the
top of the valence band (or the bottom of the conduction band) was
suggested by Hicks and Dresselhaus to improve
$ZT_{\rm
  opt}$~\cite{hicks1993effect,hicks1993thermoelectric,hicks1996experimental}.
However, we recently pointed out that in terms of their power factor
$\mathrm{PF}=S^2\sigma$, only low-dimensional semiconductors with
confinement length smaller than thermal de Broglie wavelength prove to
be useful TE materials compared with the bulk
ones~\cite{hung2016quantum}.

Another strategy to find the best thermoelectric materials is by
defining a material parameter that can be the most essential one to
determine $ZT_{\rm opt}$.  We can mention several efforts by
researchers in the past who proposed some parameters for evaluating
$ZT_{\rm opt}$.  For example, in 1996, Mahan and Sofo introduced a
dimensionless material parameter $k_B T/E_b$~\cite{mahan1996best},
where $k_B$ and $E_b$, are the Boltzmann constant and the energy band
width, respectively.  When $E_b$ is infinitesimal, the transport
distribution function $\mathcal{T}=v^2\tau\mathcal{D}$ forms a delta
function that leads to the largest possible value of $ZT_{\rm opt}$,
where $v$ is the carrier velocity, $\tau$ is the carrier relaxation
time, and $\mathcal{D}$ is the density of states of the carrier at the
Fermi energy. This work was revisited from a Landauer perspective by
Jeong et al.~\cite{jeong2012best}, they found that a finite $E_b$
dispersion produces a higher $ZT$ when the lattice thermal
conductivity is finite. Much earlier, in 1959, Chasmar and Stratton
suggested that a parameter
$B=5.745\times 10^{-6}(\mu/\kappa_l)(m/m_0)^{3/2}T^{5/2}$, where
$\mu$, $\kappa_l$, $m$, and $m_0$ are the carrier mobility, the
lattice thermal conductivity, the carrier effective mass, and the free
electron mass, respectively, determines the optimum
$ZT$~\cite{chasmar1959thermoelectric}.  Note that the product of $\mu$
and $(m/m_0)^{3/2}$ was commonly called weighted mobility.  A large
$B$ usually corresponds to a high $ZT$ value at a certain chemical
potential.  The advantage of the parameter $B$ is that to obtain a
good TE material, instead of checking all the interdependent transport
properties, one should look for a semiconductor with a high weighted
mobility and a low lattice thermal conductivity $\kappa_l$, which are
less dependent on each other.  Although $E_b$ and $B$ have been used
to guide researches in thermoelectricity for many years, it is not
possible to directly identify $ZT_{\rm opt}$ by using only these
parameters.  On the other hand, there have been a lot of efforts
dedicated to optimize the $\mathrm{PF}$, giving the optimum power
factor $\mathrm{PF}_{\rm opt}$ that can be obtained by changing the
chemical potential~\cite{goldsmid2010introduction}.  Since
$ZT_{\rm opt}$ generally occurs at a different chemical potential from
$\mathrm{PF}_{\rm opt}$, i.e.,
$ZT_{\rm opt} \neq \mathrm{PF}_{\rm opt} \kappa^{-1} T$, one always
needs to measure or estimate $ZT_{\rm opt}$ independently from
$\mathrm{PF}_{\rm opt}$ by checking again chemical potential
dependence of $ZT$.  Therefore, it should be useful for thermoelectric
applications if we can calculate $ZT_{\rm opt}$ from the information
of $\rm PF_{\rm opt}$ or other simple parameters.

In this paper, we propose that a new material parameter
$\alpha=(\mathrm{PF}_{\rm opt}/\kappa_l)T$ can be defined to directly
determine $ZT_{\rm opt}$.  Although, $ZT_{\rm opt}$ and
$\mathrm{PF}_{\rm opt}$ are generally optimized at \emph{different}
chemical potentials, the value of $ZT_{\rm opt}$ can be calculated
using an analytical formula that involves the so-called Lambert $W$
function, where $\alpha$ can be used as an input parameter.  Without
losing generality, the analytical formula for $ZT_{\rm opt}$ is
derived within the one-band model and nondegenerate semiconductor
approximation.  We will show that $ZT_{\rm opt}$ for both bulk and
low-dimensional semiconductors can be unified into a single universal
curve, which allows us to predict and understand the materials of
different dimensions that can have better $ZT_{\rm opt}$ by simply
calculating the $\alpha$ parameter.

The rest of this paper is organized as follows.  In Sec.~\ref{sec:th},
we start the derivation of some formulas of thermoelectric properties
from the conventional Boltzmann transport theory.  This initial
derivation will give us PF and $ZT$ formulas involving integrals that
must be calculated numerically.  In Sec.~\ref{sec:res}, we apply a
non-degenerate semiconductor approximation so that
$\mathrm{PF}_{\rm opt}$ and $ZT_{\rm opt}$ can be obtained
analytically, which results in the universal curve of $ZT_{\rm opt}$.
Finally, in Sec.~\ref{sec:con} we conclude the paper and give a few
perspectives for future works in the field of thermoelectricity.  We
also provide some appendixes for additional information about the
derivation of the formulas and the Lambert $W$ function.

\section{Theoretical methods}
\label{sec:th}

By solving the linearized Boltzmann equations within the one-band
model and the relaxation time approximation, three TE transport
properties are related to the transport distribution function
$\mathcal{T}(E)$ as follows:
\begin{align}
\label{eq:1}
  &\sigma=q^2\mathcal{L}_0,
    \quad S=\frac{1}{qT}\frac{\mathcal{L}_1}{\mathcal{L}_0},
    \quad \kappa_e=\frac{1}{T}\left(\mathcal{L}_2-
    \frac{\mathcal{L}_1^2}{\mathcal{L}_0}\right),
\end{align}
where $\sigma$, $S$, $\kappa_e$, are the electrical conductivity, the
Seebeck coefficient, the electronic thermal conductivity,
respectively. $\mathcal{L}_i$ is the transport integral that is
defined by~\cite{mahan1996best}
\begin{align}
\label{eq:2}
  &\mathcal{L}_i=\int\mathcal{T}(E)(E-E_F)^i
    \left(-\frac{\partial f_0}{\partial E}\right)\mathrm{d}E, \ i=0,1,2,
\end{align}
where $E$ is the energy of carrier, $f_0=1/[e^{(E-E_F)/k_BT}+1]$ is
the Fermi-Dirac distribution function, where the Fermi energy $E_F$ is
defined as the chemical potential measured from the bottom (top) of
the conduction (valence) energy band in an n-type (p-type)
semiconductor, and $\mathcal{T}(E)$ is defined
\begin{align}
\label{eq:3}
\mathcal{T}(E)=v_x^2(E)\tau(E)\mathcal{D}(E),
\end{align}
where $v_x(E)$, $\tau(E)$, and $\mathcal{D}(E)$ are the group velocity
in the $x$ direction, the relaxation time, and the density of states
(DOS) of the carrier, respectively.  

From Eqs.~\eqref{eq:1} and~\eqref{eq:2}, the thermoelectric power
factor $\mathrm{PF}$ and figure of merit $ZT$ can be written as
\begin{align}
\label{eq:4}
  \mathrm{PF} &= S^2\sigma
      = \frac{1}{T^2}\frac{\mathcal{L}_1^2}{\mathcal{L}_0}, \\
\label{eq:5}
  ZT &= \frac{S^2\sigma}{\kappa_e+\kappa_l}T
      = \beta\frac{\mathcal{L}_1^2}{\mathcal{L}_0\mathcal{L}_2-\mathcal{L}_1^2},
\end{align}
where $\kappa_l$ is the lattice thermal conductivity and
$\beta=1/(\kappa_l/\kappa_e+1)\leq 1$.  It is clear from
Eqs.~\eqref{eq:4} and~\eqref{eq:5} that the $\mathrm{PF}$ and $ZT$
have different dependence on $E_F$.

For the sake of simplicity, we consider a single parabolic band, in
which the energy band structure and the group velocity can be given as
$E(\mathbf{k})=\hbar^2\mathbf{k}^2/2m$ and
$v(\mathbf{k})=\frac{1}{\hbar}[\partial
E(\mathbf{k})/\partial\mathbf{k}]=\hbar\mathbf{k}/m$,
respectively, where $\mathbf{k}$ is the wave vector of the carrier,
$m$ is the carrier effective mass, and $\hbar$ is the Planck constant.
We assumed that the material is isotropic with a certain dimension
$d=1,2,3$, the group velocity
$v_x^2(E)=v^2(\mathbf{k})/d=\hbar^2\mathbf{k}^2/m^2d=2E/md$, and the
carrier relaxation time is inversely proportional to the carrier
DOS~\cite{zhou2011optimal}, $\tau(E)=C\mathcal{D}^{-1}(E)$, where $C$
is the scattering coefficient in units of W$^{-1}$m$^{-3}$.  The DOS
is defined as
$\mathcal{D}(E)=\frac{2}{\Omega}\sum_\mathbf{k}\delta[E-E(\mathbf{k})]$
in units of J$^{-1}$m$^{-3}$, where the factor 2 accounts for the spin
degeneracy and $\Omega$ is the volume of the system.  Detailed
derivations of how we can calculate $C$ for a typical material are
given in Appendix~\ref{app:scatt}.  After substituting $v_x^2(E)$ and
$\tau(E)$ into $\mathcal{T}(E)$ in Eq.~\eqref{eq:3}, the integrals
$\mathcal{L}_i$ in Eq.~\eqref{eq:2} can be written as
\begin{align}
\label{eq:6}
\mathcal{L}_0&=\frac{2C}{md}(k_BT)F_0, \\
\label{eq:7}
\mathcal{L}_1&=\frac{2C}{md}(k_BT)^2(2F_1-\eta F_0), \\
\label{eq:8}
\mathcal{L}_2&=\frac{2C}{md}(k_BT)^3(3F_2-4\eta F_1+\eta^2 F_0),
\end{align}
where $\eta=E_F/k_BT$ is the reduced (or dimensionless) chemical
potential and $F_j(\eta)=\int\eta^jf_0\mathrm{d}\eta$ is the
Fermi-Dirac integral.  By substituting $\mathcal{L}_i$ in
Eqs.~\eqref{eq:6},~\eqref{eq:7}, and~\eqref{eq:8} into
Eqs.~\eqref{eq:4} and~\eqref{eq:5}, we obtain the formulas of the
$ \mathrm{PF}$ and $ZT$ as follows:
\begin{align}
\label{eq:9}
   \mathrm{PF} &= \frac{2Ck_B^3T}{md}\frac{(2F_1-\eta F_0)^2}{F_0}, \\
\label{eq:10}
  ZT&=\beta\frac{(2F_1-\eta F_0)^2}{F_0(3F_2-4\eta F_1+\eta^2 F_0)
      -(2F_1-\eta F_0)^2},
\end{align}
where the integrals $F_0$, $F_1$, and $F_2$ are calculated
numerically.

\begin{figure}[t!]
  \centering \includegraphics[clip,width=8.5cm]{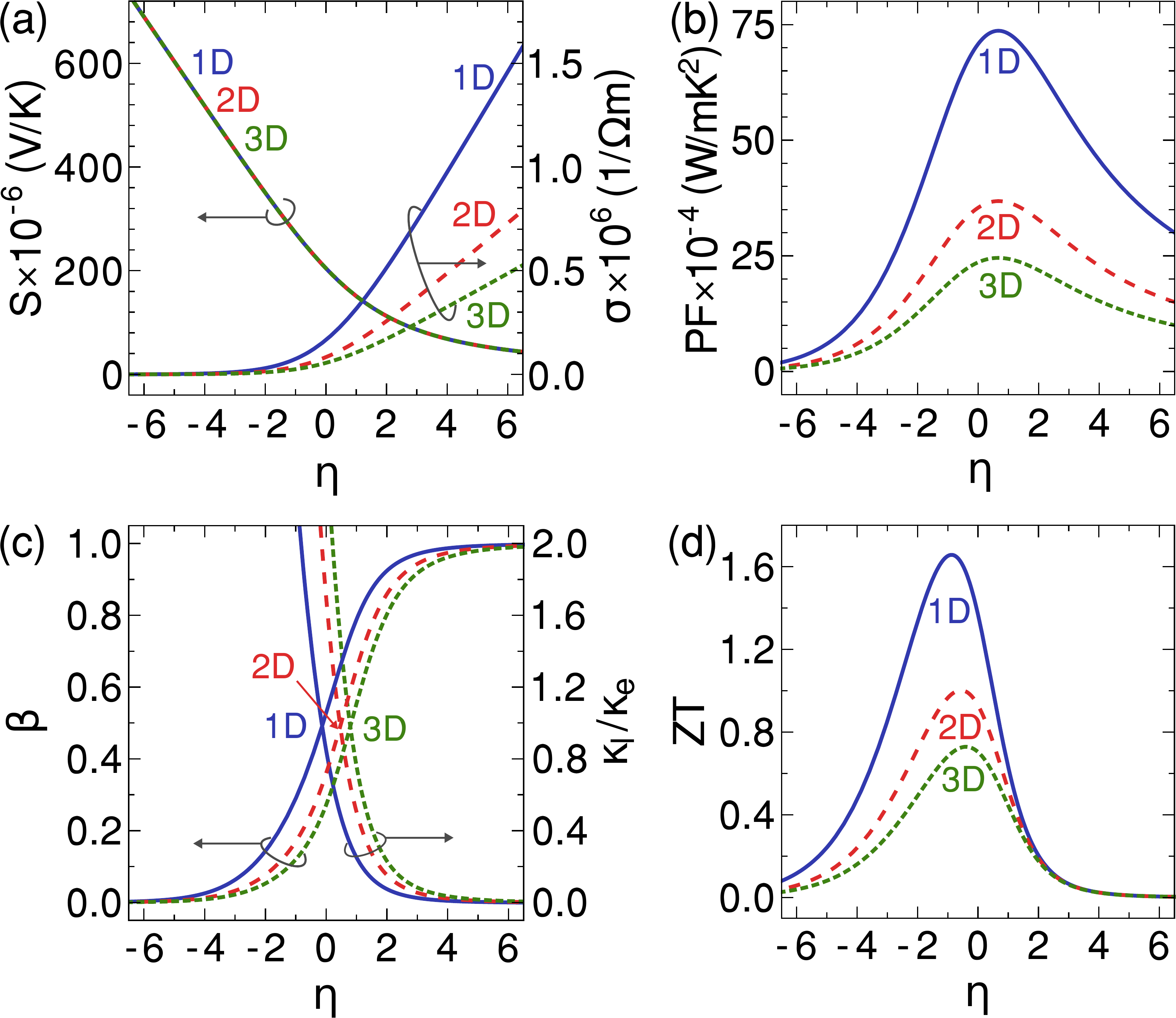}
  \caption{\label{fig:1}(a) $S$ and $\sigma$, (b) $ \mathrm{PF}$, (c)
    $\beta$ and $\kappa_l/\kappa_e$, and (d) $ZT$ as a function of the
    reduced chemical potential $\eta$ for the 1D, 2D, and 3D systems,
    respectively. The carrier effective mass, the carrier mobility,
    and the lattice thermal conductivity are set to be $m=1.12m_0$,
    $\mu=173$ cm$^2$/Vs, and $\kappa_l=0.728$ W/mK, respectively, for
    n-type Bi$_2$Te$_{2.7}$Se$_{0.3}$ at room temperature ($T=298$
    K)~\cite{liu2011thermoelectric}.}
\end{figure}

\section{Results and discussion}
\label{sec:res}

In this section we firstly discuss an example of calculating the PF
and $ZT$ as a function of $\eta$ for one semiconducting material by
using Eqs.~\eqref{eq:9} and~\eqref{eq:10} numerically.  After that, we
simplify the PF and $ZT$ formulas by considering nondegenerate
semiconductor approximation, which gives us analytical formulas of
$\textrm{PF}_{\textrm{opt}}$ and $ZT_{\textrm{opt}}$.  The
$ZT_{\textrm{opt}}$ formula can then be plotted and compared with
various experimental data, leading to a universal curve of
$ZT_{\textrm{opt}}$.

\subsection{Example of a typical material}
Figures~\ref{fig:1}(a)-(d) show, respectively, the dependence of $S$
and $\sigma$, the $ \mathrm{PF}$ [Eq.~\eqref{eq:9}], $\beta$ and
$\kappa_l/\kappa_e$, and $ZT$ [Eq.~\eqref{eq:10}] on the reduced
chemical potential $\eta$ for different dimensions.  When plotting
Figs~\ref{fig:1}(a)-(d), we consider a typical semiconductor, n-type
Bi$_2$Te$_{2.7}$Se$_{0.3}$, at $T=298$~K and the doping concentration
about $0.92\times 10^{19}$ cm$^{-3}$.  The carrier effective mass,
carrier mobility, lattice thermal conductivity are taken to be
$m=1.12m_0$, $\mu=173$ cm$^2$/Vs, and $\kappa_l=0.728$ W/mK,
respectively, for the 3D ($d=3$) bulk n-type
Bi$_2$Te$_{2.7}$Se$_{0.3}$~\cite{liu2011thermoelectric}.  The
scattering coefficient $C=1.18\times 10^{33}$ W$^{-1}$m$^{-3}$ is
obtained from $m$ and $\mu$ by using Eq.~\eqref{eq:S13} in
Appendix~\ref{app:scatt}, which leads to an average relaxation time of
about $0.1$~ps.  We temporarily use the same parameter values of $m$,
$\kappa_l$, and $C$ for the 1D ($d=1$) and 2D ($d=2$) systems as the
3D's.  However, these parameters generally vary by dimensions for
different materials, as we adopt later in Sec.~\ref{sec:univ}.

Figure~\ref{fig:1}(a) shows that $S$ is independent of $d$ and it
increases with decreasing $\eta$, while $\sigma$ depends on $d$ and it
decreases with decreasing $\eta$.  This behavior can be understood in
terms of their units since the units [V/K] of $S$ show no dependence
of length scale, while the unit [1/$\Omega$m] of $\sigma$ show
dependence of length scale.  Figure~\ref{fig:1}(b) shows a strong
enhancement of the maximum $\mathrm{PF}$ around $\eta\approx 0$ in the
low-dimensional systems (1D and 2D).  For the bulk (3D) system, the
theoretical maximum $\mathrm{PF}$ value is about $0.0025$~W/mK$^2$,
which is in a good agreement with the experimental data of about
$0.0021$~W/mK$^2$~\cite{liu2011thermoelectric}.  In the case of
$\eta \gg 0$, we can see that $S$ approaches zero because the system
becomes metallic at high doping concentrations, while $\sigma$ is
close to zero when $\eta \ll 0$ [Fig.~\ref{fig:1}(a)].  Therefore, the
$\mathrm{PF}_{\rm opt}$ occurs at $\eta\approx 0$, in which $E_F$ lies
at the bottom (top) of conduction (valence) energy band in a p-type
(n-type) semiconductor, for all the 1D, 2D, and 3D systems, as shown
in Fig.~\ref{fig:1}(b).  Figure~\ref{fig:1}(d) shows a strong
enhancement of the maximum $ZT$ values in the 1D and 2D systems, which
is known as the Hicks-Dresselhaus
theory~\cite{hicks1993effect,hicks1993thermoelectric}.  For the 3D
system, the theoretical maximum $ZT$ value is about $0.72$, which is
in a good agreement with the experimental data of about
$0.73$~\cite{liu2011thermoelectric}.  In the case of $\eta \gg 0$, the
coefficient $\beta=1/(\kappa_l/\kappa_e+1)\approx1$ since $k_e$ is
much larger than $k_l$ when the system is metallic, as shown in
Fig.~\ref{fig:1}(c).  In contrast, $\beta\approx0$ when $\eta \ll 0$
because $k_e$ is near zero (few free electron carriers in the
insulators) [see Fig.~\ref{fig:1}(c)].  Therefore, $ZT_{\rm opt}$ is found at $\eta < 0$, in which $E_F$ lies in the
energy gap, as shown in Fig.~\ref{fig:1}(d).  Important information in
Figs.~\ref{fig:1}(b) and (d) is that the $\mathrm{PF}$ and $ZT$ are
optimized at $\eta \approx 0$ and $\eta < 0$, respectively, for all
1D, 2D, and 3D systems, although the two quantities are located at
\emph{different} $\eta$ for each $d$.

\begin{figure}[t!]
  \centering \includegraphics[clip,width=8.5cm]{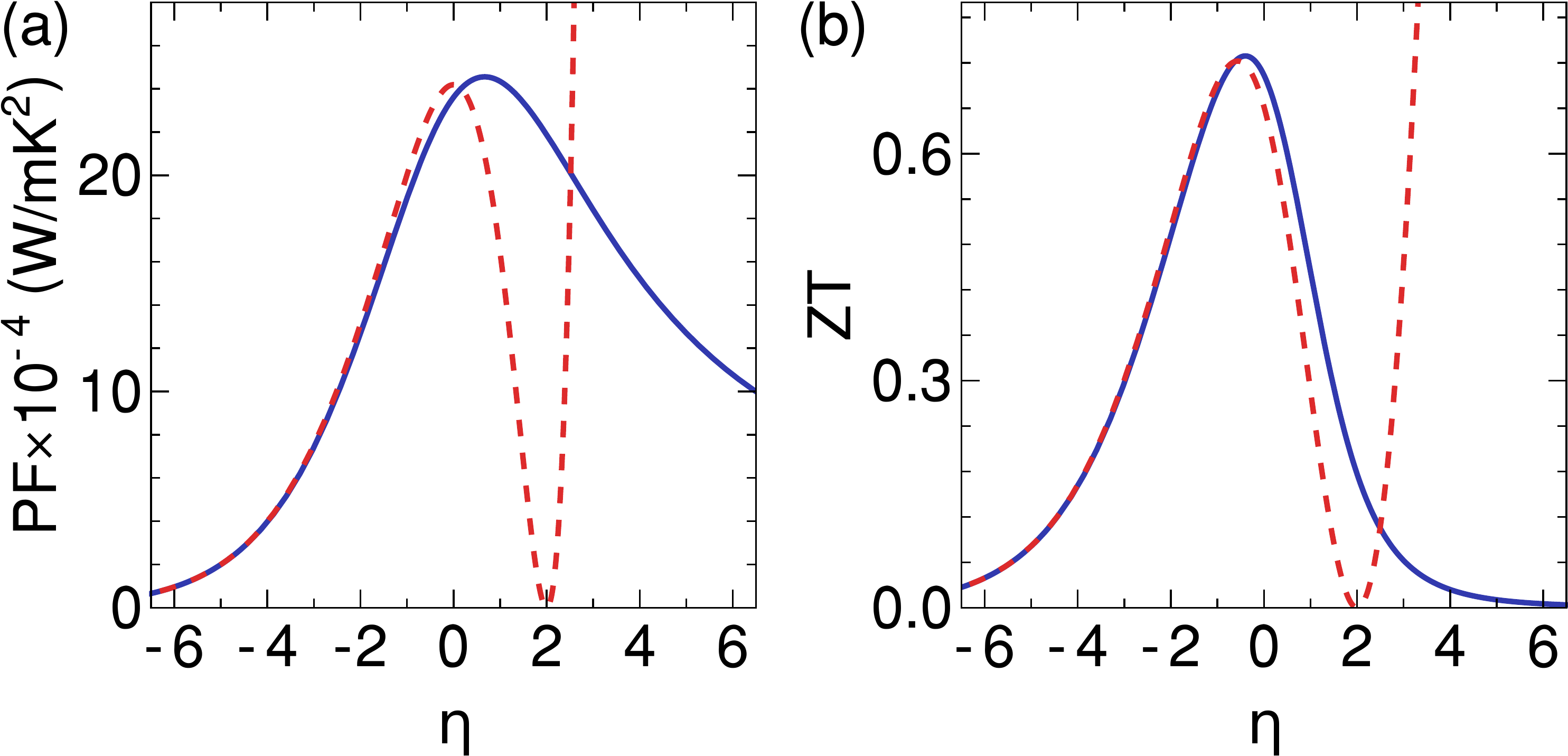}
  \caption{\label{fig:2}(a) PF and (b) $ZT$ as functions of the
    reduced chemical potential $\eta$.  Results from the formulas
    involving numerical integrations and those from analytical
    calculation (nondegenerate semiconductor approximation) are
    represented by solid and dashed lines, respectively.  The carrier
    effective mass, the carrier mobility, and the lattice thermal
    conductivity are set to be $m=1.12m_0$, $\mu=173$ cm$^2$/Vs, and
    $\kappa_l=0.728$ W/mK, respectively, for 3D n-type
    Bi$_2$Te$_{2.7}$Se$_{0.3}$ at room
    temperature~\cite{liu2011thermoelectric}.}
\end{figure}

\subsection{Nondegenerate semiconductor approximation}

Next, we would like to obtain the analytical formulas for both the
$\mathrm{PF}_{\rm opt}$ and $ZT_{\rm opt}$.  In Eqs.~\eqref{eq:9}
and~\eqref{eq:10}, which were used to plot Figs.~\ref{fig:1}(b)
and~\ref{fig:1}(d), we have considered the full solutions of
Fermi-Dirac integrals $F_0$, $F_1$, and $F_2$ numerically.  The
problem is how can we get analytical formulas for
$\mathrm{PF}_{\rm opt}$ and $ZT_{\rm opt}$ to approach these two
quantities?  Since $\mathrm{PF}_{\rm opt}$ ($ZT_{\rm opt}$) is
optimized at $\eta \approx 0$ ($\eta < 0$), we may use the
nondegenerate semiconductor approximation that is especially valid for
$\eta \leq 0$~\cite{hung2015diameter}.  In this case, the Fermi-Dirac
integral is approximated as
$F_j(\eta)\approx e^\eta \Gamma(j+1)$~\cite{hung2015diameter}, where
$\Gamma(j)$ is the Gamma function.  By substituting $F_0=e^\eta$,
$F_1=e^\eta$, and $F_2=2e^\eta$ into Eq.~\eqref{eq:9}, we get the
$\mathrm{PF}$ formula as
\begin{align}
\label{eq:11}
  \mathrm{PF}=\frac{2Ck_B^3T}{md}(2-\eta)^2e^\eta.
\end{align}
Since
$\kappa_e=\frac{1}{T}(\mathcal{L}_2-\mathcal{L}_1^2/\mathcal{L}_0)
={4Ck_B^3T^2}e^\eta/({md})$
[see Eq.~\eqref{eq:1}], $\beta$ can be written as
\begin{equation}
\label{eq:12}
\beta = \frac{1}{[2/(\alpha e^\eta)] + 1},
\end{equation}
where
\begin{equation}
\label{eq:13}
\alpha = \frac{8Ck_B^3T^2}{md\kappa_l}
\end{equation}
is a dimensionless parameter.  Substituting $\beta$ into
Eq.~\eqref{eq:10} and applying the approximation of $F_j$, we obtain
\begin{align}
\label{eq:14}
  ZT = \frac{(2-\eta)^2}{[4/(\alpha e^\eta)]+2}.
\end{align}

In Figs.~\ref{fig:2}(a) and~\ref{fig:2}(b), we respectively show
$\mathrm{PF}_{\mathrm{opt}}$ and $ZT_{\mathrm{opt}}$ that are
calculated based on the full solutions of Fermi-Dirac integrals
[Eqs.~\eqref{eq:9} and~\eqref{eq:10}] and the nondegenerate
semiconductor approximation [Eqs.~\eqref{eq:11} and \eqref{eq:14}].
If we just focus on the \emph{values} of $\mathrm{PF}_{\mathrm{opt}}$
and $ZT_{\mathrm{opt}}$ (local maxima of PF and $ZT$) at
$\eta \leq 0$, we can see that the analytical formulas based on the
nondegenerate semiconductor approximation can nicely reproduce the
$\mathrm{PF}_{\mathrm{opt}}$ and $ZT_{\mathrm{opt}}$ of the full
solutions.  Therefore, we can determine the
$\mathrm{PF}_{\mathrm{opt}}$ and $ZT_{\mathrm{opt}}$ from
Eqs.~\eqref{eq:11} and~\eqref{eq:14} by solving
$\mathrm{d}(\mathrm{PF})/\mathrm{d}\eta=0$ and
$\mathrm{d}(ZT)/\mathrm{d}\eta=0$, respectively.  The formulas
obtained for $\mathrm{PF}_{\mathrm{opt}}$ and $ZT_{\mathrm{opt}}$ are
\begin{align}
\label{eq:15}
  \mathrm{PF}_{\mathrm{opt}}=\frac{8Ck_B^3T}{md}, 
  \quad ZT_{\mathrm{opt}}=\frac{W_0^2(\alpha)}{2}+W_0(\alpha),
\end{align}
where $W_0(\alpha)$ is the principal branch of the Lambert $W$
function~(see Appendix~\ref{app:W}).  By substituting the
$\mathrm{PF}_{\mathrm{opt}}$ in Eq.~\eqref{eq:15} into
Eq.~\eqref{eq:13}, the $\alpha$ parameter is now expressed in terms of the
$\mathrm{PF}_{\mathrm{opt}}$ and $\kappa_l$,
\begin{equation}
\label{eq:16}
\alpha = \frac{\mathrm{PF}_{\mathrm{opt}}}{\kappa_l}T .
\end{equation}
The corresponding reduced chemical potentials for the
$\mathrm{PF}_{\mathrm{opt}}$ and $ZT_{\mathrm{opt}}$ are
$\eta^{\mathrm{PF}}_{\mathrm{opt}}=0$ and
$\eta^{ZT}_{\mathrm{opt}}=-W_0(\alpha)$, respectively [see
Fig.~\ref{fig:2}].  Based on the simple analytical formulas in
Eq.~\eqref{eq:15}, the values of the $\mathrm{PF}_{\mathrm{opt}}$ and
$ZT_{\mathrm{opt}}$ can be calculated directly from $C$, $d$, $m$,
$\kappa_l$, and $T$, which could be measured in experiments.  For
example, in the case of 3D n-type Bi$_2$Te$_{2.7}$Se$_{0.3}$ at room
temperature, taken from Ref.~\cite{liu2011thermoelectric}, we have
$C=1.18\times 10^{33}$~W$^{-1}$m$^{-3}$~(see also
Appendix~\ref{app:scatt}), $d=3$, $m=1.12m_0$, $\kappa_l=0.728$~W/mK,
and hence $\mathrm{PF}_{\mathrm{opt}}=0.0024$~W/mK$^2$ and
$ZT_{\mathrm{opt}}=0.72$.  This analytical result agrees well with
both fully numerical calculation
($\mathrm{PF}_{\mathrm{opt}}=0.0025$~W/mK$^2$ and
$ZT_{\mathrm{opt}}=0.72$) [see Fig.~\ref{fig:2}] and experimental data
($\mathrm{PF}_{\mathrm{opt}}=0.0021$~W/mK$^2$ and
$ZT_{\mathrm{opt}}=0.73$)~\cite{liu2011thermoelectric}.

To gain insight into the $\mathrm{PF}_{\mathrm{opt}}$, we can
substitute the coefficient $C$ in Eq.~\eqref{eq:S13} from
Appendix~\ref{app:scatt} to the $\mathrm{PF}_{\mathrm{opt}}$ formula
in Eq.~\eqref{eq:15}, so that the $\mathrm{PF}_{\mathrm{opt}}$ is
given by
\begin{align}
\label{eq:17}
  \mathrm{PF}_{\mathrm{opt}}=\frac{16\mu k_B^2}{qL^3} 
  \left(\frac{L}{\Lambda}\right)^d
  \frac{\Gamma\left(\frac{5}{2}\right)}{\Gamma\left(\frac{7-d}{2}\right)
  \Gamma\left(\frac{d}{2}\right)},
\end{align}
where $L$ is the confinement length for a particular material
dimension, and $\Lambda=[2\pi\hbar^2/(mk_BT)]^{1/2}$ is the thermal de
Broglie wavelength (a measure of the thermodynamic uncertainty for the
localization of a electron or hole of mass
$m$)~\cite{kittel1973thermal}.  Equation~\eqref{eq:17} shows the
dependence of the $\mathrm{PF}_{\mathrm{opt}}$ on $\mu$, $d$, $L$, and
$\Lambda$.  By scaling the $\mathrm{PF}_{\mathrm{opt}}$ with the
optimum $\mathrm{PF}$ of a 3D system, i.e.
$\mathrm{PF}_{\mathrm{opt}}^{\mathrm{3D}}$, we find that the ratio
$\mathrm{PF}_{\mathrm{opt}}/\mathrm{PF}_{\mathrm{opt}}^{\mathrm{3D}}$
merely depends on the factor $(L/\Lambda)^{d-3}$, consistent with our
previous work~\cite{hung2016quantum}.  It is clear that the
$\mathrm{PF}_{\mathrm{opt}}$ is enhanced for 1D and 2D semiconductors
only when $L$ is smaller than $\Lambda$.  Interestingly, in this
present study, we find that by defining
$\alpha=(\mathrm{PF}_{\mathrm{opt}}/\kappa_l)T$, we can have a direct
relation of $ZT_{\mathrm{opt}}$ with $PF_{\mathrm{opt}}$ through
Eq.~\eqref{eq:15}.  Note that $W_0(\alpha)$ monotonically increases with
$\alpha$, as shown in Fig.~\ref{fig:W} in Appendix~\ref{app:W}.  It is important
to point out that the factor $(L/\Lambda)^{d-3}$ is not only the
enhancement factor of the $\mathrm{PF}_{\mathrm{opt}}$, but also of
$ZT_{\mathrm{opt}}$ for the low-dimensional semiconductors.

\begin{figure}[t!]
  \centering \includegraphics[clip,width=8.5cm]{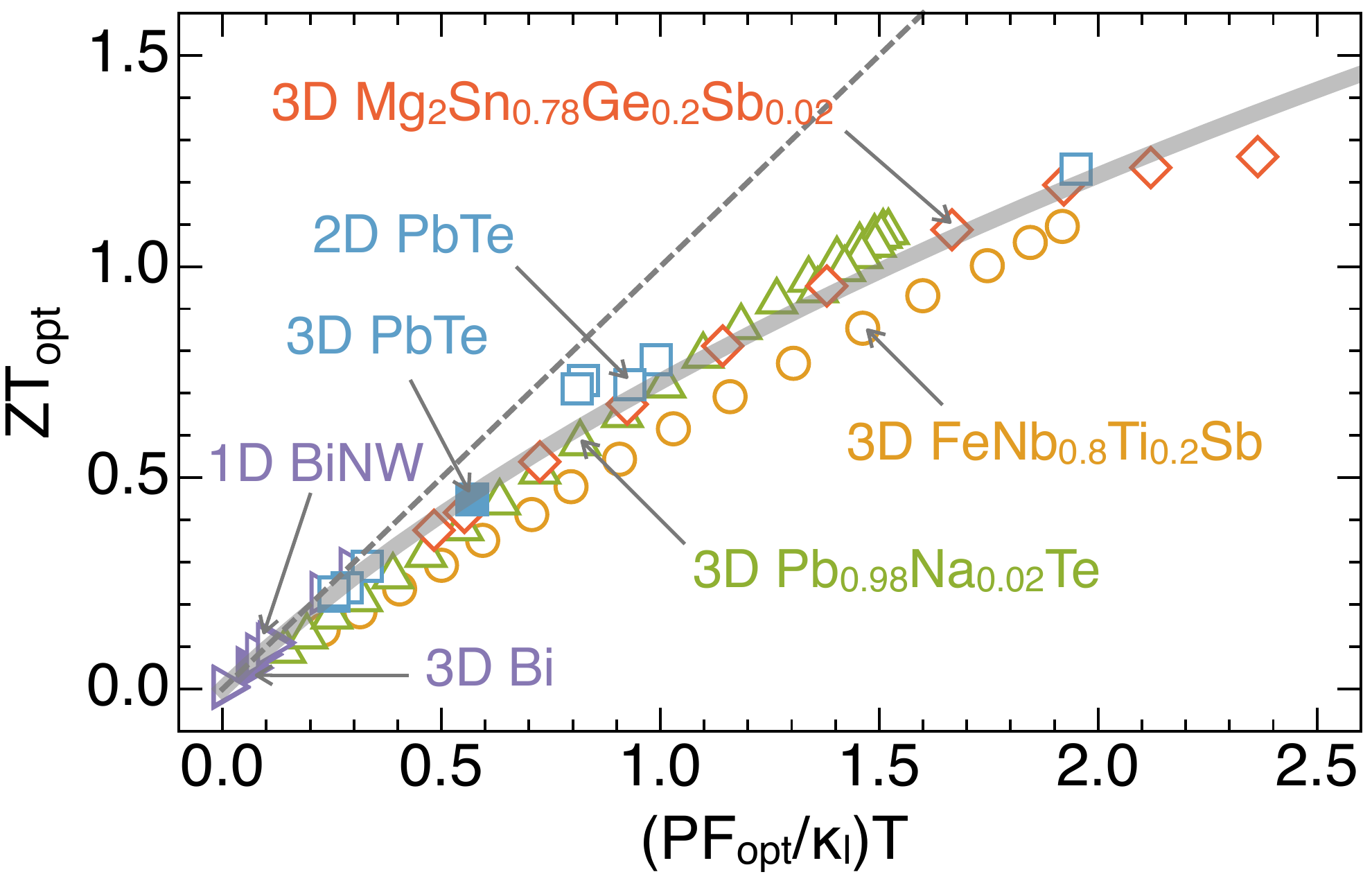}
  \caption{\label{fig:3} $ZT_{\rm opt}$ as a function of
    $\alpha=(\mathrm{PF}_{\rm opt}/\kappa_l)T$.  The solid line
    denotes the theoretical curve from Eq.~\eqref{eq:15}, while the
    dashed line is the plot of $ZT_{\rm opt} = \alpha$ as a guide for
    eyes. The symbols represent experimental results of 1D Bi
    nanowires ($\triangleright$) and 3D Bi
    ($\blacktriangleright$)~\cite{kim2015diameter}, 2D PbTe quantum
    wells ($\Box$) and 3D PbTe ($\blacksquare$)~\cite{harman1996high},
    3D Pb$_{0.98}$Na$_{0.02}$Te ($\vartriangle$)~\cite{zhao2013all},
    3D FeNb$_{0.8}$Ti$_{0.2}$Sb ($\circ$)~\cite{fu2015band}, and 3D
    Mg$_2$Sn$_{0.78}$Ge$_{0.2}$Sb$_{0.02}$
    ($\diamond$)~\cite{liu2016new}, respectively.}
\end{figure}

\subsection{The universal curve}
\label{sec:univ}

Let us now compare the $ZT_{\mathrm{opt}}$ formula with various
experimental data.  In Fig.~\ref{fig:3}, we plot theoretical
$ZT_{\mathrm{opt}}$ (solid curve) as a function of $\alpha$
[Eq.~\eqref{eq:15}].  Here $ZT_{\mathrm{opt}}$ merely depends on
$\mathrm{PF}_{\mathrm{opt}}$, $\kappa_l$, and $T$, despite of the fact
that the $\mathrm{PF}$ and $ZT$ are optimized at different chemical
potentials, i.e., $\eta^{\mathrm{PF}}_{\mathrm{opt}}=0$ and
$\eta^{ZT}_{\mathrm{opt}}=-W_0(\alpha)$, respectively.  Hence,
$ZT_{\mathrm{opt}}$ from various materials with different dimensions
can be compared directly with the theoretical curve.  The experimental
data (symbols) in Fig.~\ref{fig:3} are extracted from plots of
$ZT_{\mathrm{opt}}$, $\mathrm{PF}_{\mathrm{opt}}$, and $\kappa_l$ in
Refs.~\cite{kim2015diameter,harman1996high,fu2015band,zhao2013all,liu2016new}
by using digitizer software.  These data include 1D Bi nanowires of
different diameters ($\sim 38$--$290$~nm) along with bulk 3D Bi at
room temperature~\cite{kim2015diameter}, 2D PbTe quantum wells of
different thicknesses ($\sim 1.9$--$4.0$~nm) along with 3D PbTe at
room temperature~\cite{harman1996high}, also 3D
Pb$_{0.98}$Na$_{0.02}$Te~\cite{zhao2013all}, 3D
FeNb$_{0.8}$Ti$_{0.2}$Sb~\cite{fu2015band}, and 3D
Mg$_2$Sn$_{0.78}$Ge$_{0.2}$Sb$_{0.02}$~\cite{liu2016new} at different
temperatures ($\sim 300$--$1100$~K).

As can be seen in Fig.~\ref{fig:3}, all experimental data tend to fit
the theoretical curve from Eq.~\eqref{eq:15}.  The values of
$ZT_{\mathrm{opt}}$ monotonically increase as a function of $\alpha$ and
thus we can say that any semiconductor should have the material
parameter $\alpha>4.5$ to obtain $ZT_{\mathrm{opt}}>2$. At smaller $\alpha$
values (higher $T$ or higher $\mathrm{PF}_{\mathrm{opt}}$), we have
$\eta^{ZT}_{\mathrm{opt}}\sim\eta^{\mathrm{PF}}_{\mathrm{opt}}$,
especially around $\alpha < 0.3$.  In this case,
$ZT_{\mathrm{opt}}\sim (\mathrm{PF}_{\rm opt}/\kappa_l)T$ [see the
dotted line in Fig.~\ref{fig:3}].  On the other hand, at larger $\alpha$,
we have $\eta^{ZT}_{\mathrm{opt}}<\eta^{\mathrm{PF}}_{\mathrm{opt}}$
that eventually results in a nonlinear function of $ZT_{\mathrm{opt}}$
versus $(\mathrm{PF}_{\rm opt}/\kappa_l)T$.  The main benefit of using
the universal curve in Fig.~\ref{fig:3} is that it provides a new way
to directly calculate $ZT_{\mathrm{opt}}$ from
$\mathrm{PF}_{\mathrm{opt}}$ and $\kappa_l$ without any necessity to
check the electron thermal conductivity $\kappa_e$ nor the optimum
chemical potential $\eta^{ZT}_{\mathrm{opt}}$.

\section{Conclusion}
\label{sec:con}

We have shown that the simple analytical formulas [Eq.~\eqref{eq:15}]
based on the one-band model can directly relate the optimum figures of
merit $ZT_{\mathrm{opt}}$ and the optimum power factors
$\mathrm{PF}_{\rm opt}$ of semiconductors with different dimensions.
By introducing the material parameter
$\alpha=(\mathrm{PF}_{\rm opt}/\kappa_l)T$, we can obtain the universal
curve of $ZT_{\mathrm{opt}}$ combining both bulk and low-dimensional
semiconductors, in which $ZT_{\mathrm{opt}}$ monotonically increases
as a function of $\alpha$.  Since this approach reduces parameters such as
$\kappa_e$ and $\eta^{ZT}_{\mathrm{opt}}$ in the calculation of
$ZT_{\mathrm{opt}}$, we believe that it will help researchers better
identify new thermoelectric materials in the future.

\section*{Acknowledgments}
This work is dedicated to the late Prof. M. S. Dresselhaus.
N.T.H. and A.R.T.N thank Dr. E. H. Hasdeo for fruitful discussions and
acknowledge the financial support from the Interdepartmental Doctoral
Degree Program for Multi-dimensional Materials Science Leaders, Tohoku
University.  R.S. acknowledges JSPS KAKENHI Grant Numbers JP25107005
and JP25286005.

\appendix

\section{The scattering coefficient $C$}
\label{app:scatt}

\subsection{Defining $C$ from Fermi's golden rule}

Fermi's golden rule gives the scattering rate of transitions between
discrete states $|\mathbf{k}\rangle$ and $|\mathbf{k}^\prime\rangle$
as follows~\cite{ziman1972}
\begin{align}
\label{eq:S1}
  \frac{1}{\tau(\mathbf{k}\to\mathbf{k}^\prime)}
  \approx\frac{2\pi}{\hbar}|\langle\mathbf{k}^\prime|V|\mathbf{k}\rangle|^2
  \delta[E(\mathbf{k})-E(\mathbf{k}^\prime)],
\end{align}
where $\hbar$ is the Planck constant, $V$ is the perturbation
potential, $\delta$ is the Dirac-delta function, and $E$ is the energy
dispersion.  The general scattering rate is given by the product
$2\pi/\hbar$ times the square of transition matrix element square,
times a Dirac-delta function.  For the one-band model, the scattering
rate will be between states within parabolic energy band, where a
\textit{continuum} of states exist.  In this case, the final
scattering rate will be obtained by summation over all relevant
states,
\begin{align}
\label{eq:S2}
  \frac{1}{\tau(\mathbf{k})}
  &=
  \sum_{\mathbf{k}^\prime}
  \frac{1}{\tau(\mathbf{k}\to\mathbf{k}^\prime)}\notag\\
  &=\frac{2\pi}{\hbar}
  \sum_{\mathbf{k}^\prime}|\langle\mathbf{k}^\prime|V|\mathbf{k}\rangle|^2
  \delta[E(\mathbf{k})-E(\mathbf{k}^\prime)].
\end{align}

As an example, consider the scattering rate between electron states in
the conduction band due to a point scatterer in a 3D semiconductor.
Let us consider a perturbing potential as
$V(\mathbf{r})=V_0\delta(\mathbf{r})$ for short-range interactions,
where $V_0$ is constant in units of Jm$^3$.  The matrix element
between electronic states $|\mathbf{k}\rangle$ and
$|\mathbf{k}^\prime\rangle$ can be obtained as~\cite{lundstrom2000}
\begin{align}
\label{eq:S3}
  &|\langle\mathbf{k}^\prime|V_0\delta(\mathbf{r})|\mathbf{k}\rangle|
  \notag \\
  &=\int \mathrm{d}^3\mathbf{r}
    \left(\frac{e^{-i\mathbf{k}^\prime\cdot\mathbf{r}}}{\sqrt{\Omega}}\right)V_0
    \delta(\mathbf{r})
    \left(\frac{e^{+i\mathbf{k}^\prime\cdot\mathbf{r}}}{\sqrt{\Omega}}\right)
  =
    \frac{V_0}{\Omega},
\end{align}
where $\Omega$ is the volume of the system.  After substituting the
matrix element in Eq.~\eqref{eq:S3} into Eq.~\eqref{eq:S2}, the
scattering rate can be written as
\begin{align}
\label{eq:S4}
  \frac{1}{\tau(\mathbf{k})}
  =
  \frac{2\pi}{\hbar}\left(\frac{V_0}{\Omega}\right)^2
  \sum_{\mathbf{k}^\prime}\delta[E(\mathbf{k})-E(\mathbf{k}^\prime)].
\end{align}
By using the carrier density of states (DOS), defined as
$\mathcal{D}(E)=\frac{2}{\Omega}\sum_\mathbf{k}\delta[E-E(\mathbf{k})]$
in units of J$^{-1}$m$^{-3}$, where the factor 2 accounts for the spin
degeneracy, Eq.~\eqref{eq:S4} is now expressed as
\begin{align}
\label{eq:S5}
&\frac{1}{\tau(E)}=\frac{\pi V_0^2}{\hbar\Omega}\mathcal{D}(E).
\end{align}

This example shows an important result indicating that the scattering
rate for the continuum of states is in general proportional to the
DOS, while the strength of scattering increases with the square of the
scattering potential.  The carrier relaxation time $\tau(E)$ is thus
inversely proportional to the carrier DOS:
\begin{align}
\label{eq:S6}
{\tau(E)}=C\mathcal{D}^{-1}(E),
\end{align}
where $C=\hbar\Omega/(\pi V_0^2)$ is the scattering coefficient in
units of W$^{-1}$m$^{-3}$.  Note that according to Fermi's golden
rule, the coefficient $C$ can be a constant value when the matrix
element is approximately constant.

\subsection{Calculating \textit{C} from experimental data}

Here we derive a formula of the coefficient $C$ considering a
parabolic band for any semiconductor so that $C$ can be calculated
from experimental data.  The carrier relaxation time $\tau(E)$ and the
density of states $\mathcal{D}(E)$ per unit volume are, respectively,
defined by~\cite{lundstrom2000,hung2015diameter}
\begin{align}
\label{eq:S7}
\tau(E)&=\tau_0\left(\frac{E}{k_BT}\right)^r, \\
\label{eq:S8}
  \mathcal{D}(E)&=\frac{(2m/\hbar^2)^{d/2}E^{d/2-1}}{L^{3-d}2^{d-1}\pi^{d/2}\Gamma
                  \left(\frac{d}{2}\right)},
\end{align}
where $k_B$ is the Boltzmann constant, $T$ is the average absolute
temperature, $\tau_0$ is the carrier relaxation time coefficient, $r$
is a characteristic exponent, $d=1,2,3$ denotes the dimension of the
system, $m$ is the carrier effective mass, and $L$ is the confinement
length for a particular material dimension.  For a given $\tau(E)$,
the carrier mobility is defined by
\begin{align}
\label{eq:S9}
\mu=\frac{q\langle\langle \tau(E)\rangle\rangle}{m}.
\end{align}
The average relaxation time is defined by~\cite{lundstrom2000}
\begin{align}
\label{eq:S10}
  \langle\langle \tau(E) \rangle\rangle
  \equiv\frac{\langle E\tau(E)\rangle}{\langle E\rangle}=
  \tau_0\frac{\Gamma\left(\frac{5}{2}+r\right)}{\Gamma\left(\frac{5}{2}\right)},
\end{align}
where $\Gamma$ is the Gamma function.  From
Eqs.~\eqref{eq:S7},~\eqref{eq:S9}, and~\eqref{eq:S10}, the carrier
relaxation time $\tau(E)$ can be rewritten as
\begin{align}
\label{eq:S11}
  \tau(E)=\frac{\mu m \Gamma
  \left(\frac{5}{2}\right)}{q\Gamma\left(\frac{5}{2}+r\right)}
  \left(\frac{E}{k_BT}\right)^r.
\end{align}

\begin{figure}[t!]
  \centering \includegraphics[clip,width=65mm]{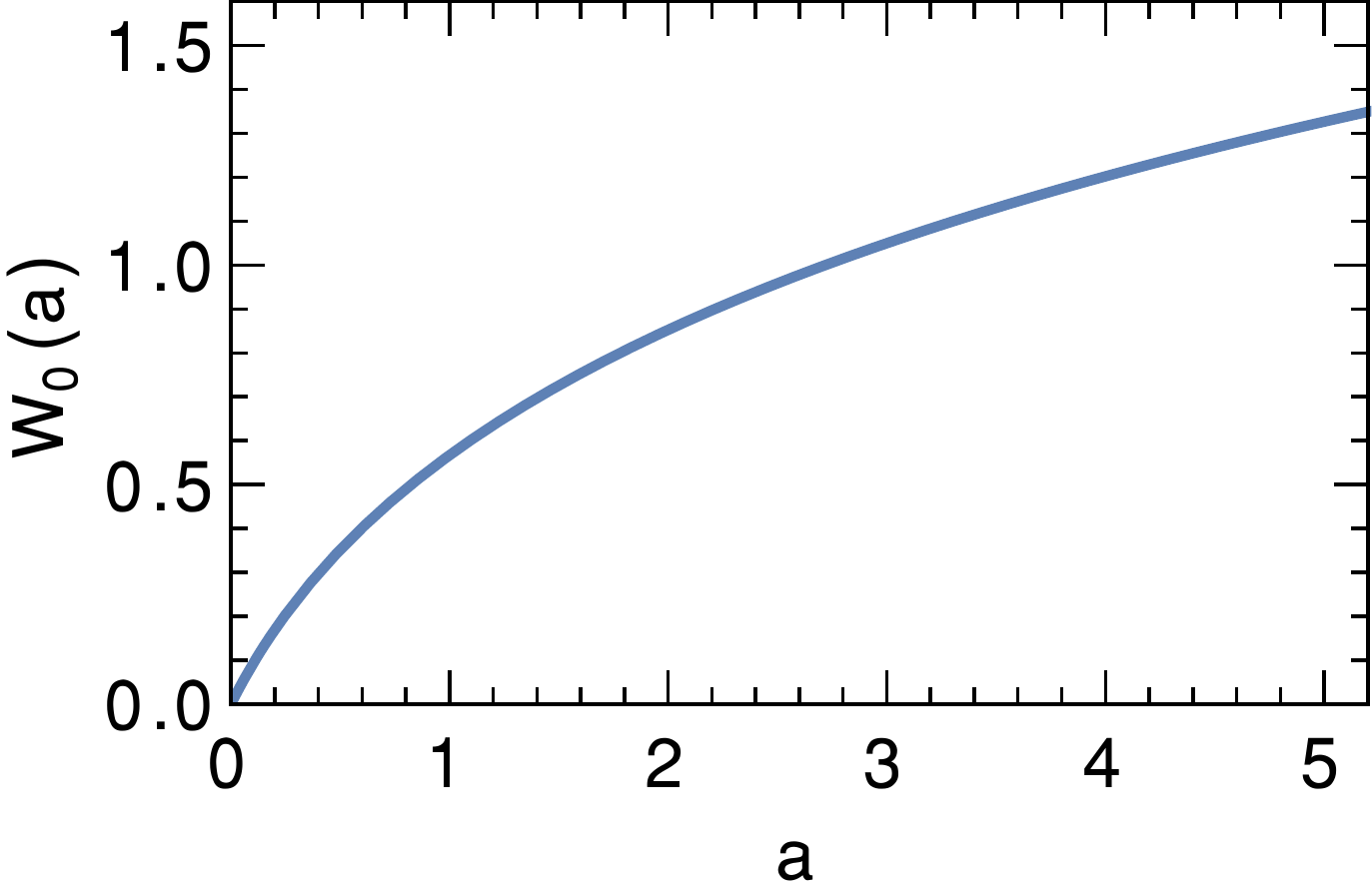}
  \caption{\label{fig:W} The real principal branch of the $W$ function
    in the case of $a\in[0,\infty)$.}
\end{figure} 

We assume that the acoustic phonon scattering is the main carrier
scattering mechanism at the room temperature, i.e.,
$\tau(E)\propto\mathcal{D}(E)^{-1}$~\cite{lundstrom2000,zhou2011optimal}.
From Eqs.~\eqref{eq:S8},~\eqref{eq:S11} and
$\tau(E)\propto\mathcal{D}(E)^{-1}$, we obtain $r=1-d/2$ for the
system with the dimension $d$.  By using $r=1-d/2$, from
Eqs.~\eqref{eq:S6},~\eqref{eq:S8}, and~\eqref{eq:S11}, the coefficient
$C$ can be written as
\begin{align}
\label{eq:S12}
  C&=\tau(E)\mathcal{D}(E)\notag\\
   &=\frac{2\mu m \Gamma
     \left(\frac{5}{2}\right)}{qk_BTL^{3-d}\Gamma
     \left(\frac{7-d}{2}\right)\Gamma\left(\frac{d}{2}\right)}
     \left(\frac{mk_BT}{2\pi\hbar^2}\right)^{d/2}.
\end{align}
After substituting the thermal de Broglie wavelength
$\Lambda=(2\pi\hbar^2/mk_BT)^{1/2}$ into Eq.~\eqref{eq:S12}, the
coefficient $C$ is given by
\begin{align}
\label{eq:S13}
  C=\frac{2\mu m}{qk_BTL^3}\left(\frac{L}{\Lambda}\right)^{d}
  \frac{\Gamma\left(\frac{5}{2}\right)}{\Gamma\left(\frac{7-d}{2}\right)
  \Gamma\left(\frac{d}{2}\right)}.
\end{align}

Equation~\eqref{eq:S13} is useful to calculate the coefficient $C$
from $\mu$ and $m$, which can be obtained from experimental data.  For
example, in the 3D ($d=3$) n-type
Bi$_2$Te$_{2.7}$Se$_{0.3}$~\cite{liu2011thermoelectric}, at room
temperature ($T = 298$ K) and doping concentration on the order of
$10^{19}$ cm$^{¡Ý3}$, the carrier mobility and the carrier effective
mass are $\mu=173$ cm$^2$/Vs and $m=1.12m_0$, respectively, where
$m_0$ is the free electron mass.  From Eq.~\eqref{eq:S13}, we obtain
the $C$ value of about $1.18\times10^{33}$ W$^{-1}$m$^{-3}$ and
correspondingly the average relaxation time is about~0.1 ps.

\section{The Lambert $W$ function}
\label{app:W}

The Lambert $W$ function is defined as a multivalued complex function that
satisfy the following equation:
\begin{align}
\label{eq:S15}
  W(\alpha) = \alpha e^{-{W(\alpha)}}, ~~~\alpha\in\mathbb{C}.
\end{align}
Equation~\eqref{eq:S15} always has an infinite number of solution in
the complex Liemann plane, hence the multivaluedness of the $W$
function.  These solutions are indexed by the integer variable $j$ and
are called the branches of the $W$ function, $W_j$, for
$j\in\mathbb{Z}$.  In particular, the solutions of Eq.~\eqref{eq:S15}
in the calculation of $ZT_{\rm opt}$ correspond to $\alpha\in[0,\infty)$.
In this case there can be a real solution, corresponding to the
principal branch of the $W$ function, i.e. $W_0(\alpha)\in[0,\infty)$.

The $W_0$ function can be written in terms of series expansion as
follows~\cite{caratheodory},
\begin{align}
\label{eq:S16}
  W_0(\alpha)
  =& \sum_{n=1}^{\infty} \frac{(-n)^{n-1}}{n!}\alpha^n \notag\\
  =& ~\alpha - \alpha^2 + \frac{3}{2}\alpha^3-\frac{8}{3}\alpha^4
     +\frac{125}{24} \alpha^5\notag\\
   &-\frac{54}{5}\alpha^6+\frac{16807}{720}\alpha^7+\cdots.
\end{align}
Figure~\ref{fig:W} shows $W_0(\alpha)$ as a function of $\alpha$ when
$\alpha\in[0,\infty)$.

%
\end{document}